\documentclass[11pt]{article}
\usepackage{latexsym}
\usepackage{graphicx}
\usepackage{amsfonts,amssymb}

\begin{document}

\title{An artificial game with equilibrium state of entangled strategy}
\author{Jinshan Wu \\
Department of Physics, Simon Fraser University, Burnaby, B.C. Canada, V5A 1S6%
}
\maketitle

\begin{abstract}
Using the representation introduced in \cite{frame}, an artificial
game in quantum strategy space is proposed and studied. Although
it has well-known classical correspondence, which has classical
mixture strategy Nash Equilibrium states, the equilibrium state of
this quantum game is an entangled strategy (operator) state of the
two players. By discovering such behavior, it partially shows the
independent meaning of the new representation. The idea of
entanglement of strategies, instead of quantum states, is
proposed, and in some sense, such entangled strategy state can be
regarded as a cooperative behavior between game players.
\end{abstract}

Key Words: Game Theory, Quantum Game Theory, Entanglement

Pacs: 02.50.Le, 03.67.-a, 03.65.Ud

\textit{Introduction} --- Recently, we proposed a new mathematical
representation for Classical and Quantum Game Theory\cite{frame}.
The idea is to define base vectors in strategy space and their
inner product so as to form them as a Hilbert space. Then a system
state is a vector in the direct product space of all single-player
state spaces. A density matrix is used to describe a system state
in this space. And then the payoff functions are reexpressed as
hermitian operators acting on this system Hilbert space. Every
player has a payoff matrix, which is a $\left(1,1\right)$-tensor
no matter how many players and how many base strategies the game
has. In that paper, many open questions were pointed out. Most of
those questions can only be discussed in the new representation.
If the conclusions of such questions shows some new phenomena, the
special value of the new representation will partially be
confirmed. One of such questions is the meaning of entangled
strategies, which will happen when the direct product relation
between system density matrix and single-player density matrix is
destroyed. In this paper, we try to answer this question by one
example. On the other hand, according to \cite{enk}, a comment on
quantum game, two questions should be answered for any quantum
version of classical game. First, whether it is helpful to solve
the original classical game; second, is a new truly quantum game,
or still in the scope of classical game. In \cite{nash}, we give a
detailed answer of those two questions, while here, we try to
provide a positive answer by this example. The conclusion in this
paper implies that from the viewpoint of classical game, besides
an equivalent description, our new representation provides an
applicable algorithm and even the process of the algorithm can be
regarded as a reasonable evolutionary process, and it could be a
way from static non-cooperative game to cooperative game; from the
viewpoint of quantum game, the game in our representation is
totally a new game, which could never be putted into the framework
of classical game, because a classical payoff matrix $G^{i}$ is
not enough to describe all information of a quantum game, which
requires a much larger matrix $H^{i}$.

\textit{The new representation} --- First, let's shortly review
the new representation. A game is defined as
\begin{equation}
\Gamma = \left(\prod_{i}^{N}\left(\times S^{i,q}\right),
\prod_{i}^{N}\left(\times S^{i,c}\right), \left\{H^{i}\right\}\right).
\label{game}
\end{equation}
in which $S^{i,q}$ has base vectors $\left\{\left|s^{i,q}_{\mu}\right>\right%
\}$, and $S^{c}_{i}$ has base vectors $\left\{\left|s^{i,c}_{\nu}\right>%
\right\}$. Usually the later is a subset of the former, but not necessary. A
classical payoff function is defined on system base vectors such as $%
H^{i,c}=\sum_{S}\left|S\right> H^{i,c}_{SS}\left<S\right|$, while
a quantum payoff function is defined as $H^{i} =
\sum_{SS^{^{\prime}}} \left|S\left>
H^{i}_{SS^{^{\prime}}}\right<S^{^{\prime}}\right|$, which has
non-zero off-diagonal elements while $H^{i,c}$ has only diagonal
terms. $H^{i}$ is hermitian and may have different forms for each
player. A system state is defined as
\begin{equation}
\rho^{s} = \prod_{i}^{N}\rho^{i}.  \label{systemstate}
\end{equation}
The payoff of player $i$ under a system state $\rho^{s}$ is
\begin{equation}
E^{i}\left(\rho^{s}\right) = Tr\left(\rho^{s}H^{i}\right).
\label{payoff}
\end{equation}
A reduced payoff matrix of player $i$, which player $i$ uses as a evaluation
of all his own strategies under the fixed strategies of all other players,
is defined as
\begin{equation}
H_{R}^{i} = Tr_{i}(\rho^{1}\cdots\rho^{i-1}\rho^{i+1}\cdots\rho^{N}H^{i}),
\label{reducedpayoff}
\end{equation}
where $Tr_{i}\left(\cdot\right)$ means to do the trace in the
space except the one of player $i$. Then from equ(\ref{payoff}),
the payoff of player $i$ also can be calculated by
\begin{equation}
E^{i}\left(\rho^{s}\right) = Tr^{i}\left(\rho^{i}H_{R}^{i}\right),
\end{equation}
in which $Tr^{i}\left(\cdot\right)$ is the trace in player $i$'s
space. An equilibrium state is defined
\begin{equation}
E^{i}\left(\rho_{eq}^{s}\right) \geq
E^{i}\left(\rho_{eq}^{1}\cdots\rho^{i}\cdots\rho_{eq}^{N}\right), \forall i.
\label{equilibrium}
\end{equation}
This definition uses the same idea as Nash Equilibrium, but has
independent meaning. First because the density matrix form allows
more strategies than the traditional mixture strategy, and second,
because $\rho_{eq}^{s}$ might be an entangled strategy state,
which destroys equ(\ref{systemstate}), while this is not allowed
in both traditional classical and quantum game. When such an
entangled state is allowed, we need to adjust a little of equ(\ref%
{equilibrium}), because at that case, $\rho_{eq}^{j}$ is not
pre-defined. Here we try to define them as reduced density matrix,
\begin{equation}
E^{i}\left(\rho_{eq}^{s}\right) \geq
E^{i}\left(Tr^{i}\left(\rho_{eq}^{s}\right)\cdot\rho^{i}\right),
\forall i. \label{extendedequilibrium}
\end{equation}
It should be noticed that when equ(\ref{systemstate}) holds, the
new definition equ(\ref{extendedequilibrium}) is equivalent with
equ(\ref{equilibrium}). A special case of the above definition is
\begin{equation}
E^{i}\left(\rho_{eq}^{s}\right) \geq E^{i}\left(\rho^{s}\right), \forall
\rho^{s}, \forall i.  \label{specialcase}
\end{equation}
Although is not always possible to find such a state
$\rho^{s}_{eq}$, in this paper, we will `produce' a game to make
use of such states. Later on, we name such equilibrium as Global
Equilibrium State (GES).

\textit{The artificial game} --- Now we define our $2$-player game on base
vector set $\left\{\left|B\right>, \left|S\right>\right\}$, which means Box
and Show respectively. We use them as base vectors for both classical and
quantum game. So the base vectors of system space are $\left\{\left|BB%
\right>, \left|BS\right>, \left|SB\right>, \left|SS\right>\right\}$. An
arbitrary system state can be
\begin{equation}
\rho^{s} = \rho_{\mu\nu}\left|\mu\right>\left<\nu\right|,
\label{generalstate}
\end{equation}
where $\mu,\nu$ is anyone of the base vectors. Now we just write
down the artificial payoff matrix,
\begin{equation}
H^{1}= \left[%
\begin{array}{cccc}
\epsilon_{1} & 0 & 0 & \epsilon_{1} \\
0 & \epsilon_{2} & \epsilon_{2} & 0 \\
0 & \epsilon_{2} & \epsilon_{2} & 0 \\
\epsilon_{1} & 0 & 0 & \epsilon_{1}%
\end{array}%
\right] = H^{2},
\label{entangledgame}
\end{equation}
which does not come from a real quantum game at this stage. But
now we will try to figure out a real quantum game with similar
situation.

The manipulative definition of a quantum game, in the traditional
framework which uses the concepts of strategy according to
\cite{jens, meyer}, has been reexpress in \cite{nash} as
\begin{equation}
\Gamma^{q,o} = \left(\rho^{q}_{0}\in\mathbb{H}^{q},
\prod_{i=1}^{N}\otimes \mathbb{H}^{i}, \mathcal{L},
\left\{P^{i}\right\}\right). \label{qogame}
\end{equation}
Where $\rho^{q}_{0}$ is the initial state of a quantum object,
$\mathbb{H}^{i}$ is player $i$'s strategy space, $\mathcal{L}$ is
a mapping from $\prod_{i=1}^{N}\otimes \mathbb{H}^{i}$ to
$\mathbb{H}^{q}$, the quantum object's operator space, and $P^{i}$
is the payoff scale for player $i$. Now in this game, the quantum
object is still chosen as spin, which has base state vectors
$\left|U\right>$ and $\left|D\right>$. In matrix form, we denote
them as $\left(1,0\right)^{T}$ and $\left(0,1\right)^{T}$. We
choose
\begin{equation}
B = \left[\begin{array}{cc}1 & 0 \\ 0 & 1\end{array}\right], S =
\left[\begin{array}{cc}0 & 1 \\ 1 & 0\end{array}\right]
\end{equation}
as our pure strategies are also the base vectors in player $i$'s
Hilbert space of quantum strategy. So a general quantum strategy
(operator) has the form,
\begin{equation}
A = xB + yS = \left[\begin{array}{cc}x & y \\ y &
x\end{array}\right].
\end{equation}
If we also require it's a unitary operator, then $A^{\dag}A = I$,
then $x$ and $y$ are not independent. The general form is
\begin{equation}
U = \cos\theta B + i\sin\theta S = \left[\begin{array}{cc}\cos\theta & i\sin\theta \\
i\sin\theta & \cos\theta\end{array}\right] \label{unitary}
\end{equation}
The initial state of the quantum object is chosen as
\begin{equation}
\rho^{q}_{0} = \left[\begin{array}{cc}\frac{3}{4} &
\frac{\sqrt{3}}{4}
\\\frac{\sqrt{3}}{4} & \frac{1}{4}
\end{array}\right].
\end{equation}
Mapping $\mathcal{L}$ is just the product $U = U^{2}U^{1}$. Payoff
scale matrix are set as
\begin{equation}
P^{1} = \frac{1}{2}\left[\begin{array}{cc}3\epsilon_{1}-\epsilon_{2} & 0 \\
0 & 3\epsilon_{2}-\epsilon_{1}\end{array}\right] = P^{2}
\end{equation}
Then in this framework, the payoff is defined as
\begin{equation}
E^{i}=Tr\left(P^{i}\mathcal{L}\rho^{q}_{0}\left(\mathcal{L}\right)^{\dag}\right)
=Tr\left(P^{i}U^{2}U^{1}\rho^{q}_{0}\left(U^{1}\right)^{\dag}\left(U^{2}\right)^{\dag}\right).
\end{equation}
This is quite similar with the quantum penny flip
game\cite{meyer}, which uses different $\rho^{q}_{0}$ and $P^{i}$.
Using the transformation procedure proposed in \cite{frame}, we
will need other two quantum pure strategies as base vectors,
$\sigma_{y}, \sigma_{z}$. So we will get a $16\times 16$-matrix as
our whole payoff matrix. And the sub-matrix related with
$\left|B\right>$ and $\left|S\right>$ is just the payoff matrix of
our artificial game. So this game should be investigated in the
larger space with the whole payoff matrix, and just because of
this, we call our game here as an artificial game. Since the
difficulty to deal with the whole payoff matrix in the larger
space, we wish the discussion of this artificial game still can
give us most information. Of course, if necessary, we can turn to
analyze the game with the whole payoff matrix. However, in this
paper, we think , a clear picture is more important than the
results of a full game.

\begin{figure}[tbp]
\includegraphics{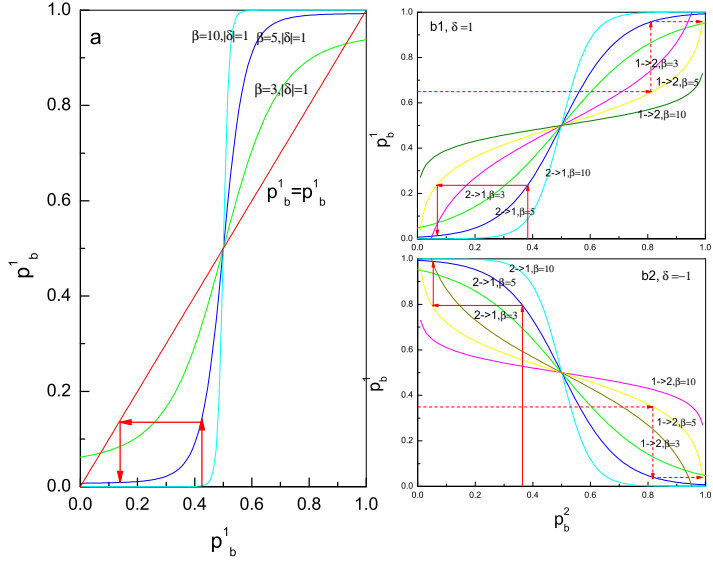}
\caption{In figure(a), self-mapping function
$p^{1}_{b}=\frac{1}{1+e^{\beta\delta\left(1-2\frac{1}{1+e^{\beta\delta\left(1-2p^{1}_{b}\right)}}\right)}}$
for different $\beta$ are plotted. For such a mapping, different
signs of $\delta$ corresponds to the same function. In figure(b),
iteration process $\left(p^{1}_{b}=
\frac{1}{1+e^{\beta\delta\left(1-2p^{2}_{b}\right)}}, p^{2}_{b}=
\frac{1}{1+e^{\beta\delta\left(1-2p^{1}_{b}\right)}}\right)$ for
different $\beta$ are plotted. From figure(a), we know that for
$p^{1}_{b}$, $0.5$ is always the unstable fixed point no matter
whether $\delta$ is positive or not and there are other two stable
fixed points depending on $\beta$. In traditional
situation\cite{frame}, under infinite resolution level ($\beta =
\infty$), the fixed points are $0$ and $1$ depending on initial
value. From figure(b), much detailed information can be extracted.
In figure(b-1), when $\delta>0$, $\left(0,0\right)$ and
$\left(1,1\right)$ are the stable fixed points, while in
figure(b-2), when $\delta<0$, they are $\left(0,1\right)$ and
$\left(1,0\right)$. So we can say, besides $\left(0.5,0.5\right)$,
when $\delta>0$, we have two NEs, but when $\delta<0$, we have
other two NEs. They are just the NEs from traditional analysis,
and here they are given by our pseudo-dynamics process.
Furthermore, thinking about the iteration process, step by step,
in some sense, this process can be regarded as an evolution. So
even the pseudo-dynamics process will have a good meaning besides
as expected it does end at the reasonable states --- NEs if they
are stable. So will it also be a way to Evolutionary Game from
Static Game?} \label{classicalplot}
\end{figure}

\textit{The classical correspondence and its solution} --- First,
we study this game in classical strategy space, which has the
general state as
\begin{equation}
\rho^{c} = \left(p^{1}_{b}\left|B\right>\left<B\right| +
p^{1}_{s}\left|S\right>\left<S\right|\right)\left(p^{2}_{b}\left|B\right>%
\left<B\right| + p^{2}_{s}\left|S\right>\left<S\right|\right),
\end{equation}
or in matrix form
\begin{equation}
\rho^{c} = \left[\begin{array}{cccc}p^{1}_{b}p^{2}_{b} & 0 & 0 &
0\\0 & p^{1}_{b}p^{2}_{s} & 0 & 0\\0 & 0 & p^{1}_{s}p^{2}_{b} &
0\\0 & 0 & 0 & p^{1}_{s}p^{2}_{s}
\end{array}\right].
\end{equation}
For such diagonal density matrix, according to the trace operator
in equ(\ref{payoff}), only the diagonal term of payoff matrix will
effect the payoff value. So payoff matrix of the classical
correspondence of this artificial game is
\begin{equation}
\begin{array}{ccc}
H^{1}= \left[%
\begin{array}{cccc}
\epsilon_{1} & 0 & 0 & 0 \\
0 & \epsilon_{2} & 0 & 0 \\
0 & 0 & \epsilon_{2} & 0 \\
0 & 0 & 0 & \epsilon_{1}%
\end{array}%
\right] = H^{2}, & \mbox{or traditionally,} & G= \left[%
\begin{array}{cc}
\epsilon_{1},\epsilon_{1} & \epsilon_{2},\epsilon_{2} \\
\epsilon_{2},\epsilon_{2} & \epsilon_{1},\epsilon_{1}%
\end{array}%
\right].%
\end{array}
\label{classicalpayoff}
\end{equation}
They are similar with Battle of the Sexes
($\epsilon_{1}>\epsilon_{2}$) and Hawk-Dove
($\epsilon_{1}<\epsilon_{2}$). The solution for a general mixture
strategy can be solved by the pseudo-dynamical way introduced by
\cite{frame}. The general reduced payoff matrix are
\begin{equation}
\begin{array}{ccc}
H^{1}_{R} = \left[%
\begin{array}{cc}
p^{2}_{b}\epsilon_{1}+p^{2}_{s}\epsilon_{2} & 0 \\
0 & p^{2}_{b}\epsilon_{2}+p^{2}_{s}\epsilon_{1}%
\end{array}%
\right] & \mbox{and} & H^{2}_{R} = \left[%
\begin{array}{cc}
p^{1}_{b}\epsilon_{1}+p^{1}_{s}\epsilon_{2} & 0 \\
0 & p^{1}_{b}\epsilon_{2}+p^{1}_{s}\epsilon_{1}%
\end{array}%
\right].%
\end{array}%
\end{equation}
So the iteration equation given by the pseudo-dynamical equation
is
\begin{equation}
p^{i}_{b}=\frac{1}{1+e^{\beta\delta\left(1-2p^{\left(3-i\right)}_{b}\right)}},
\end{equation}
in which $\delta = \epsilon_{1}-\epsilon_{2}$. The fixed points of
this iteration is shown in fig(\ref{classicalplot}).

\textit{Solution in quantum pure strategy space} --- If a general
unitary operator as equ(\ref{unitary}) can be used as the strategy
of player $1$ and player $2$, what the equilibrium state? In order
to simplify our discussion, unlike in classical game we deal with
both cases of $\epsilon_{1}>\epsilon_{2}$ and
$\epsilon_{1}<\epsilon_{2}$, here we only focus on the former
case. A general state of player $i$ is
$U^{i}\left(\theta^{i}\right)$ in equ(\ref{unitary}). Then the
reduced payoff matrix when player $2$ chooses
$\left|U^{2}\left(\theta^{2}\right)\right>$ is
\begin{equation}
H^{1}_{R} = Tr_{1}\left(\rho^{2}H^{1}\right) =
\left[\begin{array}{cc}\epsilon_{1}\cos^{2}\theta^{2}+\epsilon_{2}\sin^{2}\theta^{2}
& i\left(\epsilon_{1}- \epsilon_{2}\right)
\cos\theta^{2}\sin\theta^{2}\\i\left(\epsilon_{2}-
\epsilon_{1}\right) \cos\theta^{2}\sin\theta^{2} &
\epsilon_{2}\cos^{2}\theta^{2}+\epsilon_{1}\sin^{2}\theta^{2}
\end{array}\right].
\end{equation}
When $\epsilon_{1}>\epsilon_{2}$ the eigen-state with maximum
eigenvalue is
\begin{equation}
E^{1} = \epsilon_{1}, \left|\epsilon_{1}\right> =
\left(\cos\theta^2, -i\sin\theta^{2}\right)^{T}
\end{equation}
Compared with equ(\ref{unitary}), we know
\begin{equation}
\theta^{1} = -\theta^{2}. \label{relation21}
\end{equation}
On the other hand, if we solve the inverse question that the
solution of player $2$ when player $1$'s strategy is fixed at
$U^{1}\left(\theta^{1}\right)$, we can get
\begin{equation}
\theta^{2} = -\theta^{1}. \label{relation12}
\end{equation}
The combination of equ(\ref{relation21}) and
equ(\ref{relation12}), the equilibrium state of this game in
quantum strategy space is
\begin{equation}
\begin{array}{ccc}\left(U^{1}, U^{2}\right)
 = \left(U^{1}\left(\theta\right),U^{2}\left(-\theta\right)\right), \forall \theta & \mbox{and} &
\left(E^{1}, E^{2}\right) = \left(\epsilon_{1},
\epsilon_{1}\right)
\end{array}.
\end{equation}
Discovering all solutions of a quantum game in quantum mixture
strategy space is not a trivial problem, but will not be a topic
of this paper, because here, we just want to compare these
solutions with the entangled strategy solution in next section,
not a general way to calculate all the solutions. Anyway, for this
game, since it has many pure strategy NEs, any mixture combination
of all the pure NEs will be mixture NE.

\textit{Entangled quantum game and the GES} --- Now we try to
solve the equilibrium state in the most wider strategy space,
system strategy space, or sometimes, entangled strategy space, in
which a general state can be equ(\ref{generalstate}). As in
\cite{frame}, we named it as Entangled Quantum Game because the
strategy space permits a state without direct product relation
equ(\ref{systemstate}). In fact, because in this game, both
classical and quantum game use the same base vectors, it also can
be named as Entangled Classical Game according to the rule we
proposed in \cite{frame}. The payoff matrix of
equ(\ref{entangledgame}) have Global Equilibrium State (GES). The
payoff matrix can be rewritten as
\begin{equation}
H^{1} = H^{2} = \epsilon_{1}\left(\left|BB\right> +
\left|SS\right>\right)\left(\left<BB\right| + \left<SS\right|\right) +
\epsilon_{2}\left(\left|BS\right> +
\left|SB\right>\right)\left(\left<BS\right| + \left<SB\right|\right)
\end{equation}
So the eigen-state with maximum eigenvalue is $\left|BB\right> +
\left|SS\right>$ when $\epsilon_{1}$ is bigger and is $\left|BS\right> +
\left|SB\right>$ when $\epsilon_{2}$ is bigger. And it's easy to know they
are GES when $\epsilon_{1}$ or $\epsilon_{1}$ is larger respectively. So the
equilibrium state is
\begin{equation}
\left\{
\begin{array}{ccc}
\rho^{S}_{ges} = \left|BB\right>\left<BB\right| +
\left|BB\right>\left<SS\right| + \left|SS\right>\left<BB\right| +
\left|SS\right>\left<SS\right| & E^{2} = E^{1} = 2\epsilon_{1}&
\left(\epsilon_{1}>\epsilon_{2}\right) \\
\rho^{S}_{ges} = \left|BS\right>\left<BS\right| +
\left|BS\right>\left<SB\right| + \left|SB\right>\left<BS\right| +
\left|SB\right>\left<SB\right|& E^{2} = E^{1} = 2\epsilon_{2}&
\left(\epsilon_{1}<\epsilon_{2}\right)
\end{array}
\right.
\end{equation}
Both of these states are entangled states between the players.
This implies that an entangled strategy can win over both quantum
and classical players. And even more, since it's GES,
\begin{equation}
E^{i}\left(\rho^{S}_{ges}\right)>E^{i}\left(Tr^{1}\left(\rho^{S}_{ges}\right)Tr^{2}\left(\rho^{S}_{ges}\right)\right).
\end{equation}
This means when we destroy the correlation between player $1$ and
player $2$, both players get less payoff. In some sense, this
implies that those two players should negotiate and reach
agreement. This is the topic of Cooperative Game. So, could we
generally say, if an entangled state in system space has the
property that
\begin{equation}
E^{i}\left(\rho^{S}\right)>E^{i}\left(Tr^{1}\left(\rho^{S}\right)Tr^{2}\left(\rho^{S}\right)\right),
\forall i,
\end{equation}
it will imply a cooperative behavior?

\textit{Discussion} --- From the results above, we know solution
in quantum strategy space includes solution in classical strategy
space as special case, while solution in entangled strategy space
beats quantum solution. At the first sight of the equilibrium
entangled strategy, one might regard it as a natural result of the
classical correspondence. Because, in the case of
$\epsilon_{1}>\epsilon_{2}$, in some sense, the entangled
equilibrium means they choose to stay together
($\frac{1}{2}\left(\left|BB\right> +
\left|SS\right>\right)\left(\left<BB\right| +
\left<SS\right|\right)$). But the NEs of the classical
correspondence, equ(\ref{classicalpayoff}), are $\left|BB\right>\left<BB%
\right|, \left|SS\right>\left<SS\right|$ and
$\frac{1}{2}\left|BB\right>\left<BB\right|+\frac{1}{2}
\left|SS\right>\left<SS\right|$. They are different with our GES.
So it's totally a new phenomena in entangled game. However,
although it seems correct theoretically, how to experimentally
entangle two operators, not the usual meaning as entanglement of
quantum objects? Another question is if there is no GES in the
game, how to find the equilibrium state defined in
equ(\ref{extendedequilibrium}). In \cite{frame}, we proposed a
pseudo-dynamical iteration process on the basis of Kinetics
Equation in Statistical Mechanics, and use it to calculate the
equilibrium state of classical game. It seems work well, although
we are still pursuing a general proof. But still, we have no
applicable algorithm for quantum game. Is it possible to
generalize this approach into quantum game? At last, the same
payoff matrix of player $1$ and player $2$ makes the classical
correspondence of our artificial game not very like a battle, so
it's exactly a quantized version of Battle of the Sexes and
Hawk-Dove Game. This is also one of the reasons that we call this
game and artificial game. Anyway, according to the manipulative
definition of the quantum game $\Gamma^{q,o}$ in
equ(\ref{qogame}), it still can be realized by quantum object and
operators.

\textit{Acknowledgement} --- The authors want to thank Dr.
Shouyong Pei and Zengru Di for their advices during the revision
of this paper. This work is partial supported by China NSF
70371072 and 70371073.


\begin{thebibliography}{9}
\bibitem{frame} Jinshan Wu, A new mathematical representation of Game
Theory, arXiv:quant-ph/0404159.

\bibitem{enk} S. J. van Enk and R. Pike, Classical rules in quantum
games, Phys. Rev. A {\bf{66}}(2002), 024306.

\bibitem{meyer}D.A. Meyer, Quantum Strategies, Phys. Rev. Lett. {\bf{82}}(1999),
1052.

\bibitem{jens} J. Eisert, M. Wilkens, and M. Lewenstein,
Quantum Games and Quantum Strategies, Phys. Rev. Lett,
{\bf{83}}(1999), 3077.

\bibitem{nash}Jinshan Wu, A new mathematical representation of Game Theory II, arXiv:quant-ph/0405183.

\bibitem{marinatto}L. Marinatto and T. Weber, A quantum approach to static games of complete
information, Phys. Lett. A {\bf{272}}(2000), 291.

\bibitem{du}J. Du, etc., Remark On Quantum Battle of The Sexes
Game, arXiv:quant-ph/0103004.

\end{thebibliography}
\end{document}